# Negative Differential Conductance and Hot Phonons in Suspended Nanotube Molecular Wires

Eric Pop,[†1,2] David Mann,[†1] Jien Cao,[†1] Qian Wang,[1] Kenneth Goodson[2] and Hongjie Dai[1,*]

[1] *Department of Chemistry and Laboratory for Advanced Materials, Stanford University, Stanford, CA 94305, USA*

[2] *Department of Mechanical Engineering and Thermal Sciences, Stanford University, Stanford, CA 94305, USA*

Freely suspended metallic single-wall carbon nanotubes (SWNTs) exhibit reduced current carrying ability compared to those lying on substrates, and striking negative differential conductance (NDC) at low electric fields. Theoretical analysis reveals significant self-heating effects including electron scattering by hot non-equilibrium optical phonons. Electron transport characteristics under strong self-heating are exploited for the first time to probe the thermal conductivity of individual SWNTs ($\sim 3600$ $Wm^{-1}K^{-1}$ at $T$=300 K) up to ~700 K, and reveal a $1/T$ dependence expected for Umklapp phonon scattering at high temperatures.

* E-mail: hdai@stanford.edu

[†] *These authors contributed equally.*

The current carrying ability of materials is fundamental and important to a wide range of applications from power transmission to high performance electronics. It is now well established that when in contact with a substrate, metallic SWNTs can carry tens of micro-amperes of current.[1,2] Nevertheless, little is known about the high bias transport properties of suspended SWNTs in native unperturbed states. In general, high field transport in quasi one-dimensional (1D) materials is significantly affected by the surrounding environment due to reduced dimensionality for thermal conduction and phonon relaxation. These are also important to high power applications of 1D materials, and neither has been carefully explored.

Here, we uncover that suspended SWNTs display drastically different electron transport and phonon scattering than those on substrates. Experimentally, suspended SWNTs with Pt electrical contacts were obtained by direct growth across pre-formed trenches of widths 0.6-10 μm (Fig.1), as described previously.[3,4] Devices comprised of an individual nanotube with a non-suspended (lying on silicon nitride) and a suspended segment over a trench were also fabricated (Fig. 1a and 1b) for comparison. The devices were characterized by atomic force microscopy (AFM) and scanning electron microscopy (SEM) to obtain nanotube diameter ($d$) and length ($L$) information. We note that the obtained suspended SWNTs are typically longer than the trench width as most tubes form across the gap at a slight angle. All electrical measurements were carried out in *vacuum* at *room temperature* (R.T., $T_0 \sim 300$ K).

The same-length suspended and non-suspended portions of a metallic SWNT ($d \sim 2.4$ nm, $L \sim 3$ μm, Fig. 2) exhibit drastically different current vs. voltage (*I-V*) characteristics at high bias (Fig. 2a). The non-suspended SWNT portion shows monotonic

*I-V* with *I* approaching ~20 μA under increasing *V*, while the current in the suspended tube reaches a peak of $I_{peak}$ ~ 5 μA followed by a pronounced NDC region (Fig. 2a). We find that current peaking and NDC are universal characteristics of all suspended SWNTs in the range *L* = 0.8-11 μm (Fig. 3a). A systematic decrease in $I_{peak}$ is seen for longer suspended tubes with $I_{peak}$~10 μA/*L* (with *L* in μm) measured across many suspended tubes (Fig. 3b). We ought to note that we have found no hysteretic behavior, and the *I-V* characteristics (with NDC) are reproducible with each repeated voltage sweep from low to high and high to low.

The NDC behavior of freely suspended SWNTs starting at electric fields as low as 200 V/cm (for *L*~ 10 μm tubes) cannot be explained by the velocity saturation expected theoretically at much higher fields (~ 5 kV/cm) under isothermal conditions.[5,6] Isothermal conditions (i.e., no appreciable self-heating) are typically assumed for electron transport in SWNTs on substrates due to heat sinking by the substrate.[1-7] The *I-V* curve can be calculated by *I=V/R(V)* with resistance $R(V) = R_c + (h/4q^2)[L + \lambda_{eff}^{RT}(V)]/\lambda_{eff}^{RT}(V)$ where $R_c$ is the contact resistance. The R.T. total electron scattering mean free path[8] (*mfp*) is $\lambda_{eff}^{RT}(V) = [1/\lambda_{ac}^{RT} + 1/\lambda_{op,em}^{RT}(V)]^{-1}$ with acoustic phonon (AC) scattering *mfp*[9,10] $\lambda_{ac}^{RT}$ ~1600 nm and optical phonon (OP) emission *mfp* $\lambda_{op,em}^{RT}(V) = \hbar\omega_{op}L/qV + \lambda_{op,min}$,[7] where the first term (*q* = electron charge) represents the distance required by electrons to reach the OP emission threshold[1] energy $\hbar\omega_{op}$ ~ 0.18 V and $\lambda_{op,min}$ ~ 15 nm[1,9] is the *mfp* of OP emission after reaching the threshold. At R.T., scattering by OP absorption can be neglected due to the large energy of optical phonons in SWNTs ($\hbar\omega_{op}$ >> $k_BT$) and their low population. The *I-V* characteristic thus calculated fits well the experimental data of the SWNT on substrate for $V < 1$V (Fig. 2a). This model overestimates currents for $V > 1$V (Fig.2a), indicating that the isothermal condition may not hold at higher biases under which self-heating starts to manifest itself in the $L \sim 3$ μm SWNT on substrate.

For suspended SWNTs in vacuum, when heat dissipation only occurs along the tube length to the contacts (1D thermal transport), we consider self-heating and a bias-dependent temperature profile of the nanotube. The *I-V* characteristics are calculated by $I=V/R(V,T)$, where $R(V,T) = R_c + (h/4q^2)[L + \lambda_{eff}(V,T)]/\lambda_{eff}(V,T)$ and

$$\lambda_{eff}(V,T) = \left(1/\lambda_{ac}(T) + 1/\lambda_{op,ems}(V,T) + 1/\lambda_{op,abs}(V,T)\right)^{-1}$$ are $V$ and in turn $T$ dependent. The AC scattering *mfp* scales as $\lambda_{ac}(T) = \lambda_{ac}^{RT}(300\text{ K}/T)$ since acoustic phonons are thermally occupied with energies $\hbar\omega_{ac} \ll k_BT$ at R.T. and above.[7] The OP emission and absorption *mfp* are $\lambda_{op,ems}(T) = \hbar\omega_{op}L/qV + \lambda_{op,\min}[N_{op}(300K)+1]/[N_{op}(T)+1]$ and $\lambda_{op,abs}(T) = \lambda_{op,\min}[N_{op}(300K)+1]/N_{op}(T)$ respectively, with $N_{op} = 1/[\exp(\hbar\omega_{op}/k_BT)-1]$ as the OP occupation number. In the limit of $N_{op} \sim 0$ (below or near R.T. for $\hbar\omega_{op} \gg k_BT$) this model gives a long OP absorption *mfp*, $\lambda_{op,abs}(T)$, as expected.

At a given bias $V$ the lattice temperature of the suspended SWNT is determined by the power dissipation $P = I^2(R-R_c)$ and the thermal conductivity $\kappa(T)$ of the tube. The average temperature along the tube is $T = T_0 + I^2(R-R_c)L/(12\kappa(T)A)$ where $T_0 \sim 300$ K is the electrode (contact) temperature, $A = \pi db$ is the cross-sectional area and $b = 0.34$ nm is the tube wall thickness. The average tube temperature is used to compute the various *mfps*, then $R(V,T)$ and the *I-V* characteristics. The SWNT thermal conductivity is expected to follow $\kappa(T)=\kappa_0(T_0/T)$ at high temperatures due to Umklapp phonon-phonon scattering[11-12] where $\kappa_0$ is the R. T. thermal conductivity. Since $R(V,T)$ and $I(V)$ depend on

$T$, which is in turn dependent on $I$ and $\kappa(T)$, we use an overdamped iterative approach to compute the $I$-$V$ characteristics until $T$ converges within 0.1 K for each bias $V$.

If the OP population is assumed in equilibrium with the lattice (AC phonon) temperature, the model (with $\kappa_0 \sim 1800$ Wm$^{-1}$K$^{-1}$) can reproduce the $I$-$V$ curve of the $L \sim 3$ μm suspended SWNT only at a staggeringly high peak lattice temperature of $T > 1500$ K at $V \sim 1.5$V. This is contradicted by the fact that suspended SWNTs under such bias were never oxidized or lost connection upon exposure to air or a partial pressure of oxygen (the onset of oxidation is known to occur around $T \sim 800$ K[13]) until $V > 2$-$2.5$ V. Without the nanotube at exceedingly high lattice temperature, a rational mechanism for the low current and NDC characteristic of suspended SWNTs is the existence of non-equilibrium, hot OPs at high bias. Since only few of the numerous optical branches of a SWNT are involved in scattering electrons[6] their density of states is relatively small. Their population will quickly build up in a suspended tube if the emitted OPs cannot immediately decay into other (lower energy or electrically inactive) modes (Fig. 2b inset) or into the contacts. This is consistent with the over-population of the radial breathing mode (RBM) of suspended SWNTs observed in recent low temperature tunneling experiments[14] and with narrower Raman phonon linewidths for suspended SWNTs.[15] Recent theory has also suggested non-equilibrium optical phonons may exist under high biases in SWNTs on substrates.[16]

We model the non-equilibrium optical phonons with an effective temperature $T_{op}^{eff} = T_{ac} + \alpha(T_{ac} - T_0)$. This linear relation can be written since Joule power is primarily dissipated to the OP modes (which is the case for high bias transport in SWNTs) and then decays into AC modes (Fig. 2b inset) and to the environment ($T_0$). This is similar to previous approaches that have been used for treating non-equilibrium phonons in GaAs and Ge.[17,18,19] The non-equilibrium coefficient $\alpha > 0$ can be interpreted as the ratio of the thermal resistance of OP-AC decay to that of AC conduction along the tube, to the contacts at $T_0$. The *I-V* curves are then calculated as $I = V/R(V, T_{op}^{eff}, T_{ac})$. We found that with $\kappa(T_{ac}) = \kappa_0(T_0/T_{ac})$ where the R.T. thermal conductivity $\kappa_0 \sim 3600$ Wm$^{-1}$K$^{-1}$ (in line with that of multi-walled nanotubes[20]) and $\alpha = 2.4$, our model reproduced the experimental *I-V* curves including NDC remarkably well over a wide range of suspended SWNT lengths and biases[21] (Fig. 2a and Fig. 3a). The self-heating ($T_{op}^{eff}$ and $T_{ac}$) of the suspended SWNTs were also calculated at various biases (Fig. 2b). Importantly, the $I \sim 1/V$ shape of the *I-V* curves in the NDC region is found to strongly reflect the temperature dependence of the thermal conductivity $\kappa(T) \sim 1/T$ that results from Umklapp phonon scattering at R.T. and above[11,12] (electron contribution negligible[22]). Alternative models for the *T*-dependence of the thermal conductivity (e.g., $\kappa(T) \sim$ constant, or $\kappa(T) \sim$ linearly decreasing in *T*) cannot reproduce the *I-V* characteristics of suspended SWNTs (Fig. 4 dashed lines). Thus, we show that self-heating and *I-V* measurements on suspended SWNTs can be exploited to probe their thermal conductivity from R.T. up to ~700 K (*V* up to ~2.2 V).

Our study suggests that SWNTs suspended in vacuum, by virtue of their free unperturbed state, present the extreme scenario of longest OP lifetime and hence the strongest non-equilibrium OP population at high bias. The higher currents observed in nanotubes on substrates are owed to the substrate-tube interaction that aids heat dissipation and more importantly assists the relaxation of OPs emitted through electron scattering. Self-heating and hot phonons are also thought to exist in nanotubes lying on substrates,[16] although at higher biases (> 1 V in Fig. 2a) and electric fields than in suspended tubes. This raises the interesting possibility that SWNTs on substrates may be engineered to deliver higher currents than previously thought possible (for a given tube length) through rational interface design for optimized heat dissipation and OP relaxation. This scenario is of considerable consequences for electronics and may have general implications for high-current applications of quasi-1D materials. The effects uncovered here could also be exploited for new device applications of suspended SWNTs.

We acknowledge valuable discussions with Ali Javey and Sanjiv Sinha. This work is supported by the MARCO MSD Focus Center.

**References:**

[1] Z. Yao, C. L. Kane, and C. Dekker, Phys. Rev. Lett. **84**, 2941 (2000).

[2] A. Javey, P. Qi, Q. Wang, et al., Proc. Nat.Acad. Sci. **101**, 13408 (2004).

[3] J. Cao, Q. Wang, D. Wang, et al., Small **1**, 138 (2005).

[4] J. Cao, Q. Wang, M. Rolandi, et al., Phys. Rev. Lett. **93**, 216803 (2004).

[5] G. Pennington and N. Goldsman, Phys. Rev. B **68**, 045426 (2003).

[6] V. Perebeinos, J. Tersoff, and P. Avouris, Phys. Rev. Lett. **94**, 086802 (2005).

[7] J. Y. Park, S. Rosenblatt, Y. Yaish, et al., Nano Letters **4**, 517 (2004).

[8] S. Datta, *Electronic Transport in Mesoscopic Systems* (University Press, Cambridge, 1995).

[9] A. Javey, J. Guo, M. Paulsson, et al., Phys. Rev. Lett. **92**, 106804 (2004).

[10] D. Mann, A. Javey, J. Kong, et al., Nano Lett. **3**, 1541 (2003).

[11] J. M. Ziman, (Oxford Univ. Press, 1960).

[12] M. A. Osman and D. Srivastava, Nanotechnology **12**, 21 (2001).

[13] I. W. Chiang, B. E. Brinson, R. E. Smalley, et al., J. Phys. Chem. B **105**, 1157 (2001).

[14] B. J. LeRoy, S. G. Lemay, J. Kong, et al., Nature **432**, 371 (2004).

[15] H. Son, Y. Hori, S. G. Chou, et al., Appl. Phys. Lett. **85**, 4744 (2004).

[16] M. Lazzeri, Piscanec, S., Mauri, F., Ferrari, A.C. and Robertson J., 2005).

[17] S. Madhavi, V. Venkataraman, J. C. Sturm, et al., Phys. Rev. B **61**, 16807 (2000).

[18] H. M. Van Driel, Phys. Rev. B **19**, 5928 (1979).

[19] A. Majumdar, K. Fushinobu, and K. Hijikata, J. of Appl. Phys. **77**, 6686 (1995).

[20] P. Kim, L. Shi, A. Majumdar, et al., Phys. Rev. Lett. **87**, 215502/1 (2001).

[21] The absolute values of $\kappa_0$ and $\alpha$ here are approximate and based on the assumption of oxidation temperature of $T\sim 800$ K for a suspended SWNT Joule-heated in air. Additional uncertainty exists due to the SWNT diameter known within approximately ~10 percent from AFM measurements on the non-suspended parts.

[22] T. Yamamoto, S. Watanabe, and K. Watanabe, Phys. Rev. Lett. **92**, 075502 (2004).

**Figure Captions**

**Figure 1:** Freely suspended nanotube devices. (a) Scanning electron microscope (SEM) image, taken at a 45 degree angle, of the non-suspended (on nitride) and suspended (over a ~ 0.5 μm deep trench, entirety of trench depth not shown in the image) $L$ ~ 2 μm nanotube segments of an individual nanotube, both segments with Pt contacts under the tube.[3,4] (b) A schematic of the device cross-section. (c) SEM image of a nanotube suspended over a trench across a ~ 10 μm Pt electrode gap. The bright features between the Pt electrodes were metal lines at the bottom of the wide trench used as local gates. The diameters of the nanotubes were measured by AFM in the range of $d$ ~ 2-3 nm. Nanotubes grown under the same condition on transmission electron microscope (TEM) grids were confirmed to be single-walled by TEM.

**Figure 2:** High-field electron transport in suspended nanotubes. (a) Current-voltage (*I-V*) characteristics of the same-length ($L$~3 μm) suspended and non-suspended portions of a SWNT ($d$~2.4 nm) at room temperature measured in vacuum. The symbols represent experimental data, the lines are calculations based on average tube temperature (similar within ~5% to that based on actual tube temperature profile and resistance integrated over the ~3 μm tube length). (b) Computed average acoustic (lattice) and effective optical phonon temperature vs. bias voltage for the suspended tube segment in (a) quantifying the degree of self-heating. The figure inset shows the energy flow in our model from electrons ("heated" by the electric field) to optical phonons and then acoustic phonons. We estimated the heat dissipation by radiation (ignored in our model) to be less than 1 percent of the power input even up to average SWNT temperatures $T$ ~ 800 K. Dissipation at the contacts ($I^2 R_c$) is estimated to be less than 5 percent at high bias.

**Figure 3:** Currents in various-length suspended SWNTs. (a) *I-V* of four suspended tubes with lengths: $L$ ~ 0.8, 2.1, 3 and 11 μm respectively. Symbols are experimental data, solid lines are calculations. The *I-V* curves are highly reproducible over many sweeps without irreversible changes to the device, indicating that the electrical characteristics are not due to irreversible contact change. The tube diameters used in the calculation (consistent with AFM) were $d$ ~ 2, 2.4, 2.4 and 3.2 nm respectively. The contact resistance $R_c$ was used as a parameter to fit the *I-V* curves at low bias. $R_c$ ~ 15 kΩ for all tubes except for $L$ = 3 μm which had $R_c$ ~ 30 kΩ. This contact resistance is much lower than the resistance along the tube under high bias and most (> 95 percent) of the power dissipation occurs along the tube length rather than at the contacts. (b) Measured peak current (symbols) for suspended SWNTs of various lengths. The peak current scales approximately as ~1/$L$ just like the thermal conductance of the suspended nanotubes, an additional indicator that this is a thermally-limited effect. The deviation from the 1/$L$ behavior is attributed to variations in diameter between the different tubes.

**Figure 4:** Electrical characteristics of a 2 μm suspended tube in vacuum (experimental data with symbols) and calculations (lines) based on three different thermal conductivity models with $\kappa_0$ =

3600 Wm$^{-1}$K$^{-1}$ at $T_0$ = 300 K. The high bias region of the *I-V* curve provides an indirect measurement of the temperature dependence of the thermal conductivity for 400 K < *T* < 700 K. Here, as in the rest of our paper, *T* refers to the lattice (AC phonon) temperature.

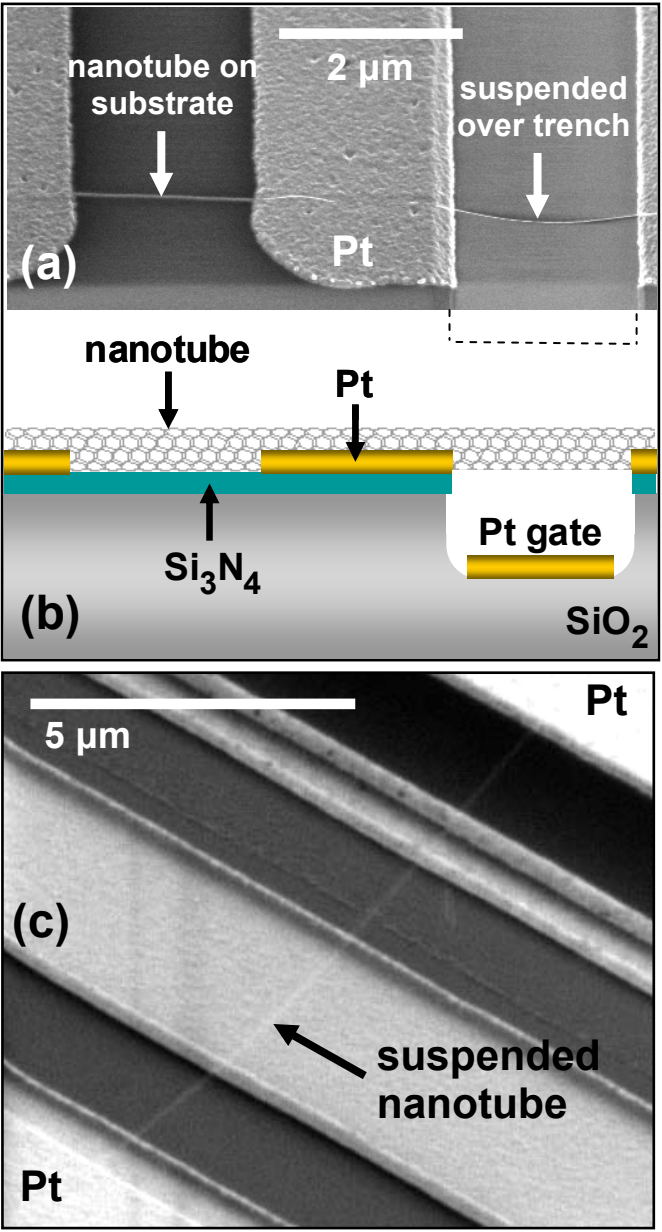

Figure 1.

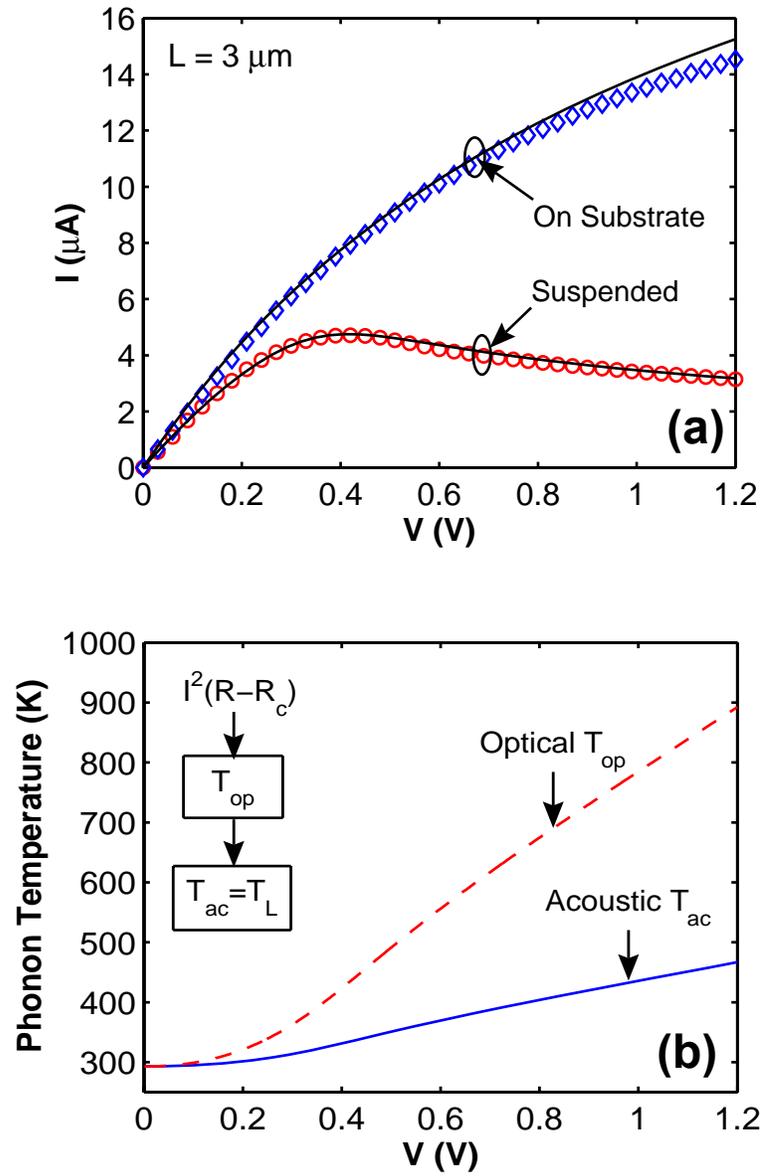

Figure 2.

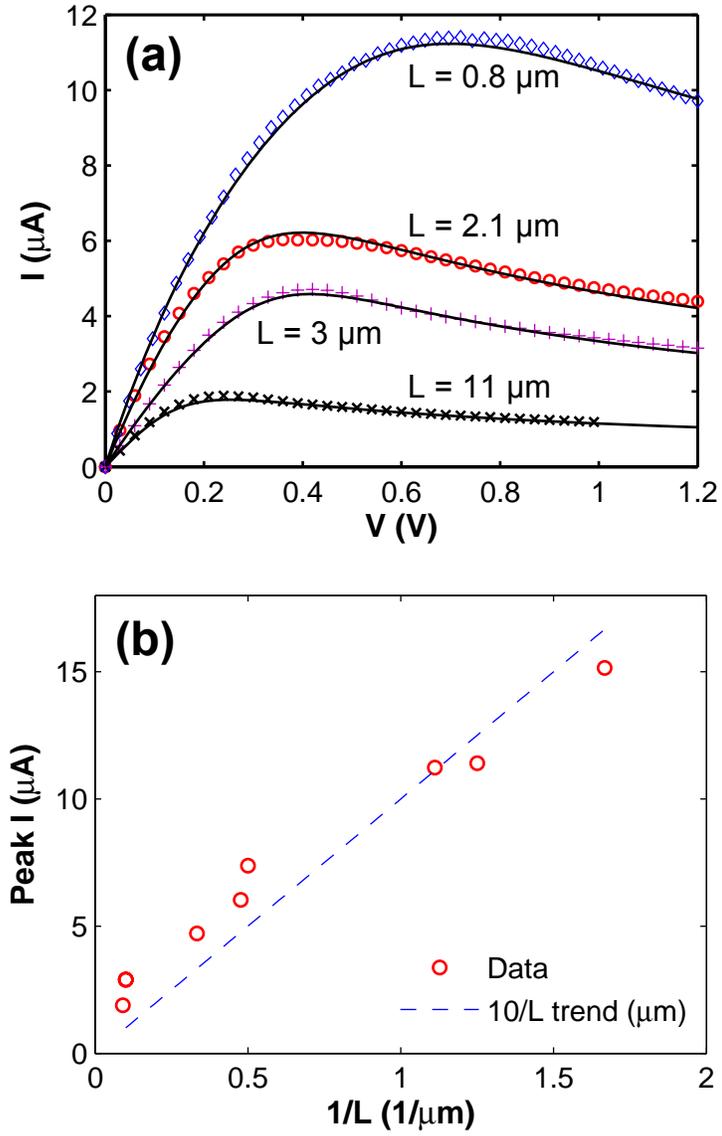

Figure 3.

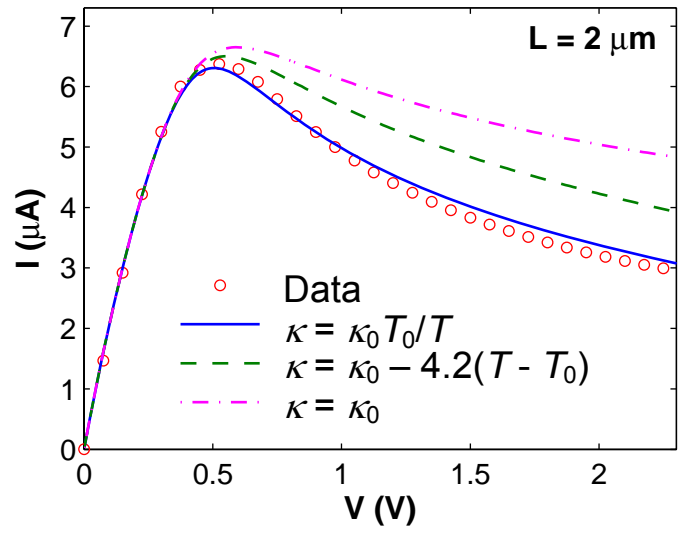

Figure 4.